\documentstyle[12pt,epsfig]{article}
\newcommand{\dd}{\mbox{{\rm d}}}

\def\lsim{\mathrel{\rlap{\raise 2.5pt \hbox{$<$}}\lower 2.5pt
\hbox{$\sim$}}}

\newcommand{\Lumint}{{\cal L}_{\rm int}}

\begin{document}
\pagestyle{empty}
\begin{titlepage}
\vspace{1. truecm}
\begin{center}
\begin{Large}
{\bf Comparative analysis of the four-fermion contact interactions at $e^+e^-$
and $e^-e^-$ colliders\footnote{Talk given at
the VII-th International School - Seminar ``The Actual Problems
of Microworld Physics'', Gomel, 28 July - 8 August, 2003}
}
\end{Large}

\vspace{2.0cm}

{\large  A.V. Tsytrinov and A.A. Pankov}
\\[0.3cm]

The Pavel Sukhoi Technical University of Gomel, Belarus

\end{center}
\vspace{1.0cm}

\begin{abstract}
\noindent
We study electron-electron contact-interaction searches in the processes
$e^+e^-\to e^+e^-,\mu^+\mu^-$, and $e^-e^-\to e^-e^-$ at planned Linear
Colliders run in the $e^+e^-$ and $e^-e^-$ modes with both beam longitudinally
polarized.

\noindent

\vspace*{3.0mm}

\noindent
\end{abstract}
\end{titlepage}
\pagestyle{plain}


Contact interaction Lagrangians (CI) provide an effective framework 
to account for the phenomenological effects of new dynamics 
characterized by extremely high intrinsic mass scales $\Lambda$, at
the `low' energies $\sqrt s\ll\Lambda$ attainable at current particle 
accelerators. For the Bhabha scattering process
\begin{equation}
e^++e^-\to e^++e^-,
\label{proc1}
\end{equation}
as well as for M{\o}ller scattering
\begin{equation}
e^-+e^-\to e^-+e^-, 
\label{proc2}
\end{equation}
we consider the flavor-diagonal, helicity conserving ${eeff}$
contact-interaction effective Lagrangian \cite{Eichten}:
\begin{equation}
{\cal L}_{\rm CI}
=\frac{1}{1+\delta_{ef}}\sum_{i,j}g^2_{\rm eff}\hskip
2pt\epsilon_{ij}
\left(\bar e_{i}\gamma_\mu e_{i}\right)
\left(\bar f_{j}\gamma^\mu f_{j}\right).
\label{lagra}
\end{equation}
In Eq.~(\ref{lagra}): $i,j={\rm L,R}$ denote left- or right-handed 
fermion helicities, $\delta_{ef}=1$ for processes (\ref{proc1}) 
and (\ref{proc2}) and, if we assumed lepton universality, the 
same Lagrangian, with $\delta_{ef}=0$, is relevant to the annihilation 
processes 
\begin{equation}
e^++e^-\to \mu^++\mu^-. 
\label{proc3}
\end{equation}
The CI coupling constants in Eq.~(\ref{lagra}) are 
parameterized in terms of corresponding mass scales as 
$\epsilon_{ij}={\eta_{ij}}/{{\Lambda^2_{ij}}}$ and, according to the previous 
remarks concerning compositeness, one assumes $g^2_{\rm eff}=4\pi$. Also, 
by convention, one takes $\vert\eta_{ij}\vert=1$ or $\eta_{ij}=0$, 
leaving the energy scales $\Lambda_{ij}$ as free, {\it a priori} independent,  
parameters.

We notice that for the case of the Bhabha process (\ref{proc1}), 
Eq.~(\ref{lagra}) envisages the existence of three independent CI models, 
each one contributing to individual helicity amplitudes or combinations of 
them, with {\it a priori} free, and nonvanishing, coefficients 
(basically, $\epsilon_{\rm LL},\epsilon_{\rm RR}\ {\rm and}
\ \epsilon_{\rm LR}=\epsilon_{\rm RL}$ combined with the $\pm$ signs). 
The same is true for the M{\o}ller process (\ref{proc2}). In 
general, apart from the $\pm$ possibility, for $e^+e^-\to{\bar f}f$ with 
$f\ne e$ there are four independent CI couplings, so that 
in the present case of processes (\ref{proc1}) and (\ref{proc2}) there is 
one free parameter less. Correspondingly, in principle, a 
model-independent analysis of the data should account for the situation 
where the full Eq.~(\ref{lagra}) is included in the expression for the 
cross section. Potentially, in this case, the different CI couplings may 
interfere and such interference could substantially weaken the bounds. 
To this aim, in the case of the processes (\ref{proc1}), (\ref{proc2}) and
(\ref{proc3}) at the Linear Collider (LC) considered here, a possibility is
offered by initial beam polarization, that enables us to extract from the data
the individual helicity cross sections (or their combinations)
through the definition of particular, polarized integrated cross
sections and, consequently, to disentangle the constraints on 
the corresponding CI constants \cite{Babich2001,Pankov2003}. 
In this note, we wish to present a model-independent analysis of the CI 
that complements that of Refs.~\cite{Pankov2003}, and is 
based instead on the measurements of more `conventional' observables 
(but still assuming polarized electron and positron beams) 
such as the differential distributions of the final leptons.
We also make a comparison of the results from these three processes.

With $P^-$ and $P^+$ the longitudinal polarization  
of the electron and positron beams, respectively, and $\theta$ the angle 
between the incoming and the outgoing electrons in the c.m.\ frame, 
the differential cross section of process (\ref{proc1}),
including $\gamma$ and $Z$ exchanges both in the $s$ and $t$ 
channels and the contact interaction (\ref{lagra}), can be written 
in the following form \cite{Pankov2003}: 
\begin{eqnarray}
\frac{\dd\sigma(P^-,P^+)}{\dd\cos\theta}
&=&\frac{(1+P^-)\,(1-P^+)}4\,\frac{\dd\sigma_{\rm R}}{\dd\cos\theta}+
\frac{(1-P^-)\,(1+P^+)}4\,\frac{\dd\sigma_{\rm L}}{\dd\cos\theta}
\nonumber \\
&+&\frac{1+P^-P^+}2\,\frac{\dd\sigma_{{\rm LR},t}}{\dd\cos\theta}.
\label{cross}
\end{eqnarray}
In Eq.~(\ref{cross}): 
\begin{eqnarray}
\frac{\dd\sigma_{\rm L}}{\dd\cos\theta}&=&
\frac{\dd\sigma_{{\rm LL}}}{\dd\cos\theta}+
\frac{\dd\sigma_{{\rm LR},s}}{\dd\cos\theta},
\nonumber \\
\frac{\dd\sigma_{\rm R}}{\dd\cos\theta}&=&
\frac{\dd\sigma_{{\rm RR}}}{\dd\cos\theta}+
\frac{\dd\sigma_{{\rm RL},s}}{\dd\cos\theta},
\label{sigP}
\end{eqnarray}
with 
\begin{eqnarray}
\frac{\dd\sigma_{{\rm LL}}}{\dd\cos\theta}&=&
\frac{2\pi\alpha^2}{s}\,\big\vert A_{{\rm LL}}\big\vert^2,
\quad
\frac{\dd\sigma_{{\rm RR}}}{\dd\cos\theta}=
\frac{2\pi\alpha^2}{s}\,\big\vert A_{{\rm RR}}\big\vert^2, 
\nonumber \\
\frac{\dd\sigma_{{\rm LR},t}}{\dd\cos\theta}&=&
\frac{2\pi\alpha^2}{s}\,\big\vert A_{{\rm LR},t}\big\vert^2, 
\quad\frac{\dd\sigma_{{\rm LR},s}}{\dd\cos\theta}=
\frac{\dd\sigma_{{\rm RL},s}}{\dd\cos\theta}=
\frac{2\pi\alpha^2}{s}\,\big\vert A_{{\rm LR},s}\big\vert^2, 
\label{helsig}
\end{eqnarray}
and
\begin{eqnarray}
A_{{\rm RR}}&=&\frac{u}{s}\,\left[1+\frac{s}{t}+g_{\rm R}^2\hskip 2pt
\left(\chi_Z(s)+\frac{s}{t}\,\chi_Z(t)\right)+
2\frac{s}{\alpha}\,\epsilon_{\rm RR}\right],
\nonumber \\
A_{{\rm LL}}&=&\frac{u}{s}\,\left[1+\frac{s}{t}+g_{\rm L}^2\hskip 2pt
\left(\chi_Z(s)+\frac{s}{t}\,\chi_Z(t)\right)+
2\frac{s}{\alpha}\,\epsilon_{\rm LL}\right],
\nonumber \\
A_{{\rm LR},s}&=&
\frac{t}{s}\,\left[1+g_{\rm R}\hskip 2pt
g_{\rm L}\,\chi_Z(s)+\frac{s}{\alpha}\,\epsilon_{\rm LR}\right],
\nonumber \\
A_{{\rm LR},t}&=&
\frac{s}{t}\,\left[1+g_{\rm R}\hskip 2pt
g_{\rm L}\chi_Z(t)+\frac{t}{\alpha}\,\epsilon_{\rm LR}\right].
\label{helamp}
\end{eqnarray}
Here: $\alpha$ is the fine structure constant; $t=-s(1-\cos\theta)/2$, 
$u=-s(1+\cos\theta)/2$ and 
$\chi_Z(s)=s/(s-M^2_Z+iM_Z\Gamma_Z)$ and  
$\chi_Z(t)=t/(t-M^2_Z)$ 
represent the $Z$ propagator in the $s$ and $t$ channels, respectively, 
with $M_Z$ and $\Gamma_Z$ the mass and width of the $Z$;  
$g_{\rm R}=\tan\theta_W$, $g_{\rm L}=-\cot{2\,\theta_W}$ are the SM 
right- and left-handed electron couplings of the $Z$, with $\theta_W$ 
the electroweak mixing angle.

With both beams polarized, the polarization of each beam can be changed 
on a pulse by pulse basis. This would allow the separate measurement 
of the polarized cross sections for each of the four polarization 
configurations RR, LL, RL and LR, corresponding to the four sets of beam
polarizations $(P^-,P^+)=(P_1,P_2)$, $(-P_1,-P_2)$, $(P_1,-P_2)$ and
$(-P_1,P_2)$, respectively, with $P_{1,2}>0$. 
To make contact to the experiment we take $P_1=0.8$ and $P_2=0.6$, and 
impose a cut in the forward and backward directions. Specifically, 
we consider the cut angular range $\vert\cos\theta\vert<0.9$ and divide 
it into nine equal-size bins of width $\Delta z=0.2$ ($z\equiv\cos\theta$).
We also introduce the experimental efficiency, $\epsilon$, for
detecting the final $e^+e^-$ pair, and according to the LEP2 experience  
$\epsilon=0.9$ is assumed. The reach on the CI couplings, and the corresponding
constraints on their allowed values in the case of no effect observed, can be
estimated by performing $\chi^2$ analysis, assuming the data to be well
described by the SM ($\epsilon_{\alpha\beta}=0$) predictions, i.e., that no
deviation is observed within the foreseen experimental uncertainty. 

The procedure, and the criteria, to derive numerical constraints from the 
M{\o}ller process and muon pair-production process (\ref{proc3}) are quite
similar. One should notice only that in the case 
of M{\o}ller scattering one can find for the cross section results similar to 
Bhabha scattering, that can be obtained by crossing symmetry except for 
the overall normalization factor 1/2 related to identical 
particles \cite{Pankov2003}. 
Also, from Eqs.~(\ref{cross})-(\ref{helamp}) one can obtain the cross 
section for muon pair-production process accounting that it proceeds 
solely via $s$-channel exchange \cite{Babich2001}.

As for the systematic uncertainty, we take $\delta\Lumint/\Lumint=0.5\%$, 
$\delta\epsilon/\epsilon=0.5\%$ and, regarding the electron and positron 
degrees of polarization, $\delta P_1/P_1=\delta P_2/P_2=0.5\ \%$.
As a criterion to constrain the values of the contact interaction
parameters allowed by the non-observation of the corresponding deviations, 
we impose $\chi^2<\chi^2_{\rm CL}$, where the actual
value of $\chi^2_{\rm CL}$ specifies the desired `confidence' level.
We take the values $\chi^2_{\rm CL}=$7.82 and 9.49 for 95\% C.L. for 
a three- (Bhabha and M{\o}ller processes) and a four-parameter ($\mu^+\mu^-$
pair production) fit, respectively.

In Figs.~1-2 we show the derived limits on the electron contact 
interactions at a LC
with longitudinally polarized beams and using a model-independent analysis
that allows to simultaneously account for all independent couplings as
non-vanishing free parameters. From these figures one can conclude that the
two processes, (\ref{proc1}) and (\ref{proc2}), are complementary as far as the
sensitivity to the individual
couplings in a model-independent data analysis is concerned: the sensitivity of
Bhabha scattering to $\Lambda_{\rm LR}$ is dramatically higher, while
M{\o}ller scattering is the most sensitive to $\Lambda_{\rm LL}$ and
$\Lambda_{\rm RR}$.

\begin{figure}[t]
 \epsfclipon
 \epsfxsize=9cm
 \centerline{
 \epsfxsize=9cm
 \epsfbox{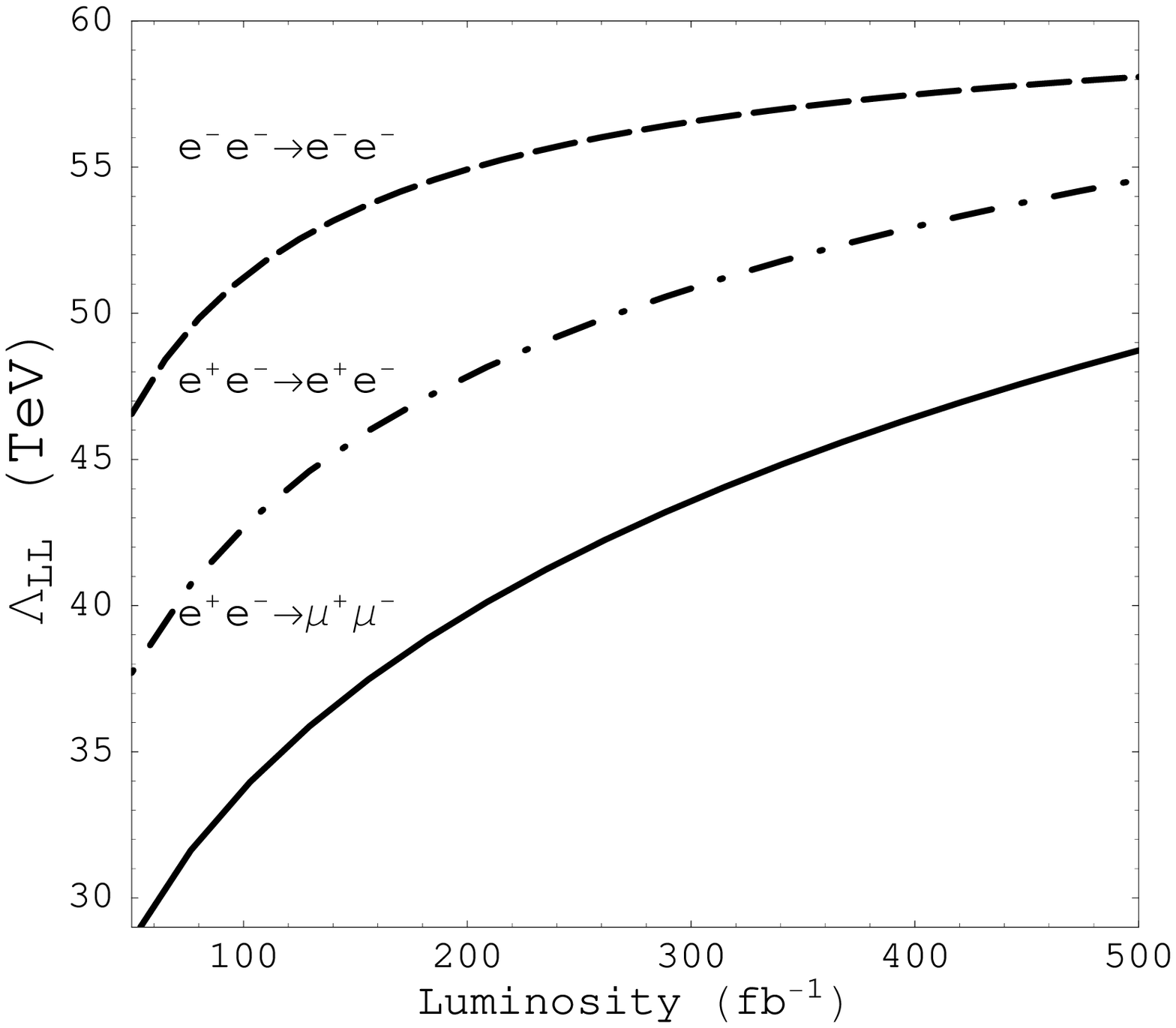}
 \epsfxsize=9cm
 \epsfbox{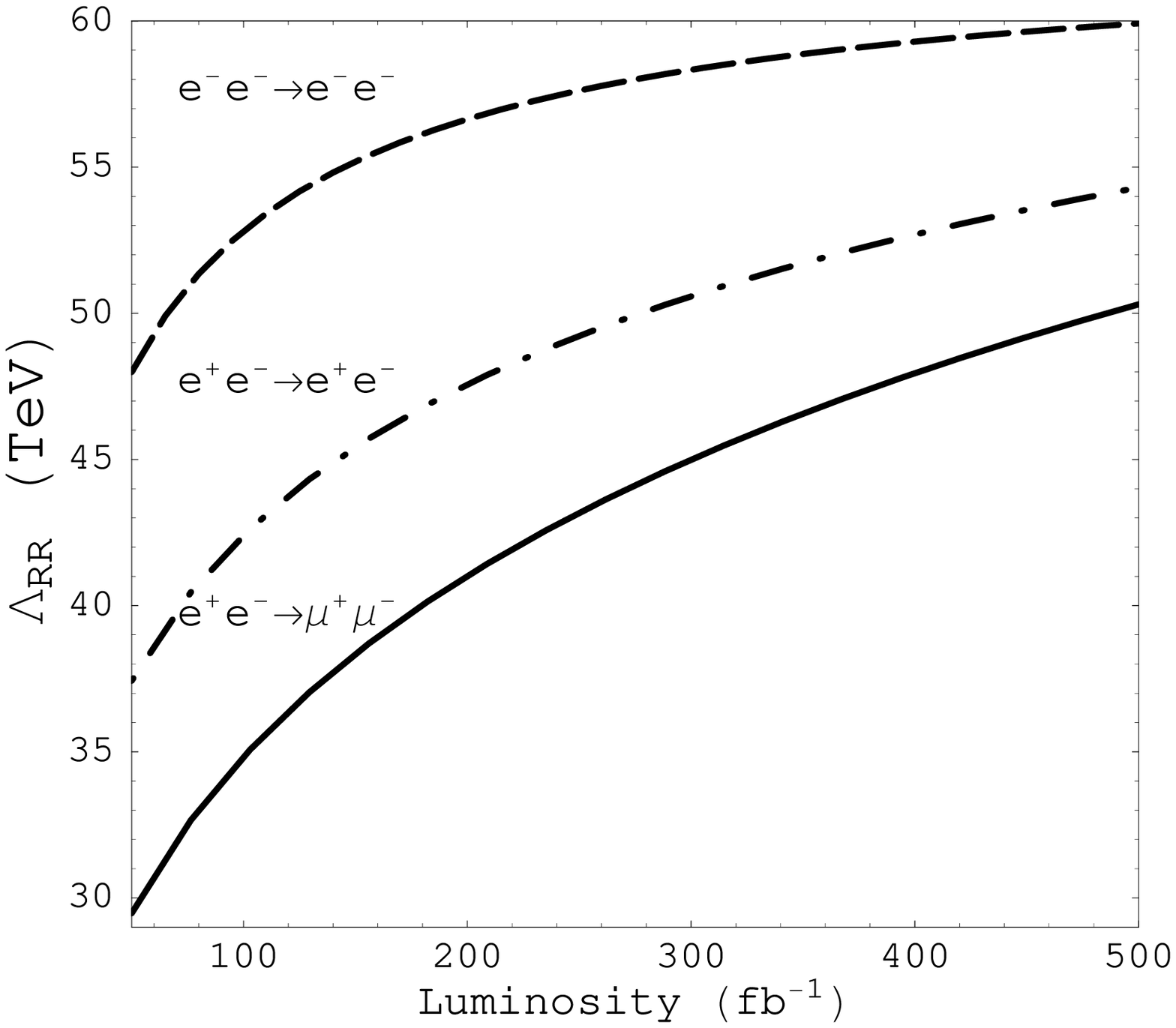}}
 \caption{Reach in $\Lambda_{\rm LL}$ and  $\Lambda_{\rm RR}$ at 95\% C.L. {\it
vs.} integrated 
luminosity ${\Lumint}$ obtained from the model-independent analysis
for $e^+e^-\to e^+e^-,\mu^+\mu^-$ and $e^-e^-\to e^-e^-$ 
at $E_{\rm c.m.}=0.5$~TeV, $\vert P^-\vert=0.8$ and $\vert P^+\vert=0.6$.}
\label{fig1}
\end{figure}

\vspace{1. truecm}
 
\begin{figure}
 \epsfxsize=9cm
\centerline{\epsfbox{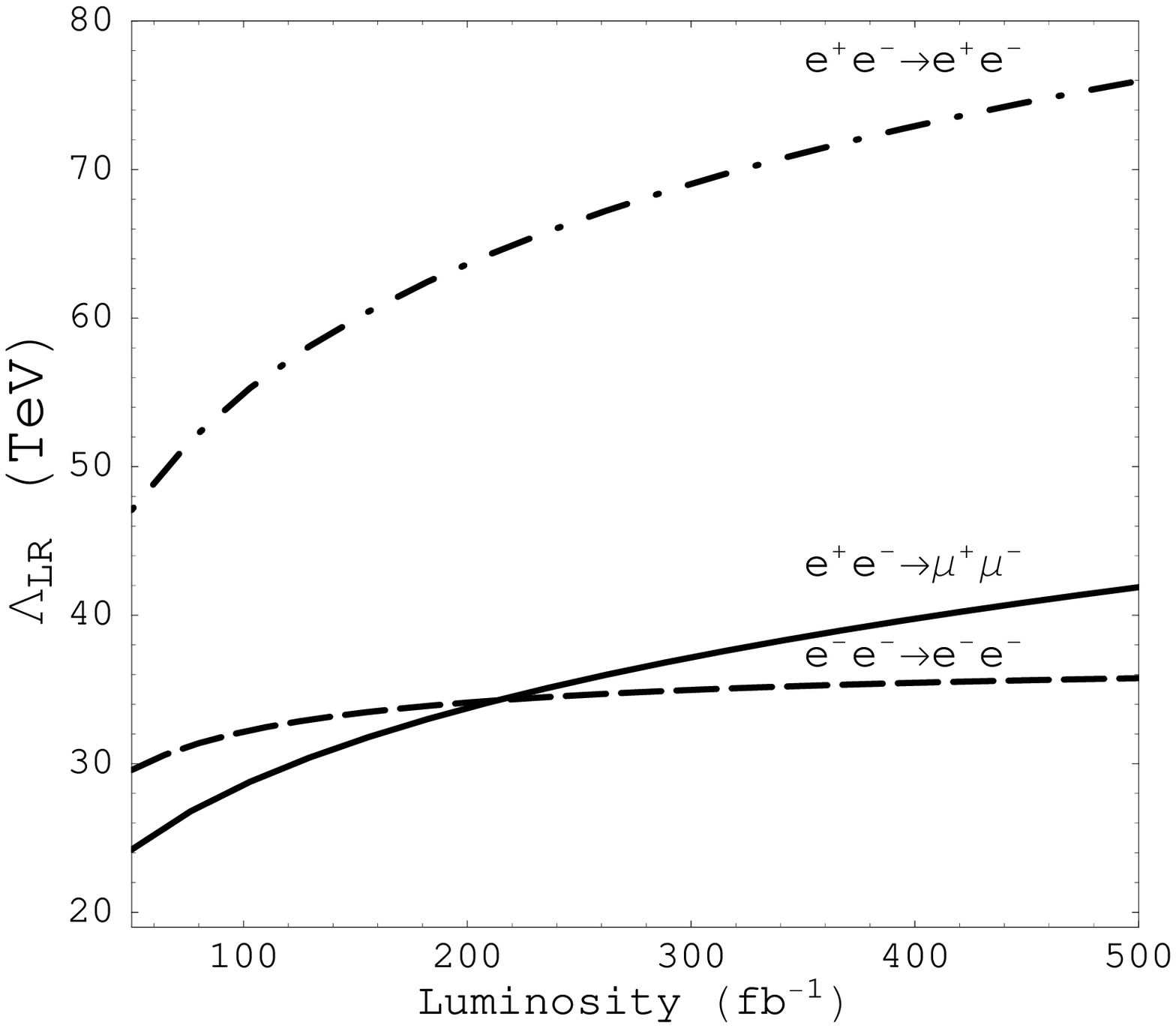}}
\caption{Same as in Fig.~1 but for $\Lambda_{\rm LR}$.
}
\label{fig3}
\end{figure}

\goodbreak


\begin{thebibliography}{99}

\bibitem{Eichten}
E.~Eichten, K.~Lane and M.~E.~Peskin,
Phys.\ Rev.\ Lett.\ {\bf 50} (1983) 811.
   
\bibitem{Babich2001} 
A.~A.~Babich, P.~Osland, A.~A.~Pankov and N.~Paver,
Phys. Lett. B {\bf 518} (2001) 128.

\bibitem{Pankov2003}
A.~A.~Pankov and N.~Paver, Eur.\ Phys.\ J.\ C {\bf 29} (2003) 313.

\end{thebibliography}
\end{document}